\documentclass[runningheads]{llncs}
\usepackage[T1]{fontenc}
\usepackage{orcidlink}  
\usepackage{enumitem}
\begin{document}
\title{Integration of AI in Cybersecurity: Current Trends with a Focused Look at Intrusion Detection Applications}
\titlerunning{Integration of AI in Cybersecurity: Current Trends in Intrusion Detection}
%
\author{Saadeddine Tazili\orcidlink{0009-0008-0890-8247} \and
Abdeljebar Mansour\orcidlink{0000-0002-5231-4736} \and
Mohamed Yassin Chkouri\orcidlink{0000-0001-7256-329X}}
\authorrunning{S. Tazili et al.}
%
\institute{SIGL Laboratory, ENSATE, Abdelmalek Essaâdi University, Tetouan, Morocco
\email{tazili.saadeddine@etu.uae.ac.ma, abdeljebar.mansour@uae.ac.ma, mychkouri@uae.ac.ma}}

\maketitle              

\begin{center}
\small
\textbf{Accepted at AI2SD 2025. Forthcoming in Springer Lecture Notes in Networks and Systems (2026).}

\textbf{Please cite this paper as:} \\
Tazili, S., Mansour, A., Chkouri, M. Y. (2026). Integration of AI in Cybersecurity: Current Trends with a Focused Look at Intrusion Detection Applications. In: Ezziyyani, M., Kacprzyk, J., Balas, V.E. (eds) International Conference on Advanced Intelligent Systems for Sustainable Development (AI2SD 2025). AI2SD 2025. Lecture Notes in Networks and Systems. Springer Nature Switzerland, Cham. To appear.
\end{center}
\vspace{0.5em}

\begin{abstract}
Artificial Intelligence (AI) is widely adopted today for its ability to detect patterns, automate tasks, and reduce time and cost across various applications. Its integration into Cybersecurity has garnered significant attention, particularly in areas such as intrusion detection, malware analysis, and phishing or spam detection. As AI and cybersecurity evolve, new methods and approaches emerge regularly. Current trends include the use of Generative AI, Natural Language Processing, Federated Learning for privacy-preserving collaborative training, and eXplainable AI to ensure interpretability and trust, which are vital in cybersecurity. This paper presents an interesting review of current AI-based cybersecurity trends, focusing on intrusion detection approaches and aiming to uncover meaningful insights through comparative analysis based on the employed AI techniques and reported performance.

\keywords{Artificial Intelligence \and AI \and Cybersecurity \and Intrusion Detection \and Generative AI \and  eXplainable AI \and XAI \and Natural Language Processing \and NLP \and Federated Learning \and FL \and Genetic Algorithms.}
\end{abstract}
\section{Introduction}
\label{sec:intro}

Artificial Intelligence (AI) is increasingly being applied in various domains due to its ability to detect patterns, automate tasks, and save time. These strengths make it valuable in Cybersecurity, both offensively, such as generating malicious code or conducting social engineering, and defensively, including intrusion detection, spam detection, malware analysis, and phishing prevention.

As AI evolves, its use in cybersecurity follows, with new trends such as the application of Generative AI and Natural Language Processing (NLP) techniques, Federated Learning for privacy-preserving collaborative training, and eXplainable AI (XAI) to ensure interpretability and trust, which are essential in a sensitive field like cybersecurity.

Several previous surveys have addressed AI in cybersecurity. For instance, Capuano et al. \cite{Capuano2022} conducted a comprehensive survey concerning the role of AI explainability in cybersecurity. They provided a structured taxonomy of XAI concepts and techniques and reviewed prior studies that sought to incorporate interpretability into AI-driven security systems. The authors asserted the extreme need for explainability in cybersecurity, emphasizing that many AI models exhibit substantial opacity despite achieving high values on conventional performance metrics (e.g., accuracy, precision, F1-score, recall). Such opacity, they argued, renders the delegation of critical cybersecurity decisions to these models inherently dangerous. On the other hand, the authors also asserted that explainability in cybersecurity can be a double-edged sword, since it may facilitate new attacks as the AI models will also be explainable to the attacker, posing severe security threats. 

Ghimire and Rawat \cite{Ghimire2022} conducted a survey concerning the use of Federated Learning (FL) for cybersecurity with a special focus on  Internet of Things (IoT) environments. They reviewed research on FL models as defenses against various cyberattacks, as well as studies examining attacks targeting FL implementations, such as poisoning and reverse engineering. Additionally, they discussed approaches to addressing key challenges that affect FL performance, including high communication costs, system heterogeneity, and statistical heterogeneity. The survey also included descriptions of commonly used datasets for evaluating Machine Learning (ML) models in cybersecurity contexts, such as KDDCup99, NSL-KDD, MNIST, CICIDS 2017, and UNSW-NB15. 

Next, Bahassi et al. \cite{Bahassi2022} made a brief review of the frequently encountered cyberattacks in recent years, such as Malware, Distributed Denial of Service (DDoS), Malicious URLs, Phishing, Impersonation attacks, and Knocking Down CAPTCHA. They also gave a brief review of frequently used ML algorithms for detecting attacks, such as support vector machine, k-nearest neighbors classifier, decision tree, logistic regression, and deep learning. Additionally, Boumahdi et al. \cite{Boumahdi2025} provided us with a literature survey of feature selection methods based on information theory, probability, genetic algorithms, and meta-heuristic techniques. 

Unlike previous works, we review in our work some of the current trends in the usage of AI in cybersecurity, especially focusing on intrusion detection. Past surveys and reviews on AI integration in cybersecurity were typically either addressing general AI applications or specific AI concepts and technologies, but relevant surveys tracking current trends in this field are still missing. The papers included in this review are all Scopus-indexed and were selected based on their relevance to intrusion detection, including their recency (excluding works published before 2019) and their citation count.

The rest of this paper is organized as follows. Section \ref{sec:res} presents the literature review. However, Section \ref{sec:conc} discusses the reviewed works, while highlighting key findings and research gaps. Eventually, we conclude the paper while outlining perspectives for future research.

\section{Literature Review}
\label{sec:res}

In this section, we review the main works related to cybersecurity, considering explainability in AI, FL, Generative AI, NLP methods, and Genetic Algorithms.

\subsection{Explainability}

Concerning the explainability, Eddermoug et al. \cite{Eddermoug2023} proposed the ``\textit{klm}-based Profiling and Preventing Security Attacks ($klm$-PPSA)'' system for cloud and IoT environments, targeting attacks at the application layer. Indeed, based on their previously published patent \cite{Eddermoug2022Patent}, they used the $klm$-PPSA Dataset V 1.0 \cite{klm-PPSA-dataset2022} in this case study. Further, this system can be adjusted to other similar network/Internet-based environments. The proposal added three security factors as classification variables: $k$ (number of login attempts with an incorrect password), $l$ (number of biometric login attempts), and $m$ (number of attempts with a correct password but an invalid keystroke). To classify login attempts as safe or not, the approach employs the Regularized Class Association Rules algorithm, which mines class-association rules that link variables to classes while satisfying the appropriate support and confidence constraints. These rules are both effective for classification and inherently interpretable.

On the other hand, Le et al. \cite{Le2022} proposed a method for intrusion detection using empirically selected ensemble trees on the IoTID20 \cite{Ullah2020} and ML-based Network Intrusion Detection Systems Datasets (NF-BoT-IoT-V2 and NF-ToN-IoT-V2) \cite{Sarhan2022}. SHapley Additive exPlanations (SHAP) values were used for explainability, providing global explanations via heatmap plots and local explanations via decision plots.

Similarly, Javeed et al. \cite{Javeed2023} applied SHAP for local explainability via decision \& waterfall plots and global explainability via the summary plot. They proposed a deep learning-based model consisting of Bidirectional Long Short-Term Memory (BiLSTM) and Bidirectional Gated Recurrent Units (BiGRU) layers for intrusion detection, trained and tested on the CICIDS2017 dataset \cite{Sharafaldin2018}.  

Sharma et al. \cite{Sharma2024} proposed two intrusion detection architectures -- a Deep Neural Network (DNN) and a three-block Convolutional Neural Network (CNN) -- both using ReLU activations. The DNN trained faster and was therefore selected for explainability analysis with Local Interpretable Model-agnostic Explanations for local explanations and SHAP for local and global explanations. Both models were evaluated on the NSL-KDD \cite{Tavallaee2009} and UNSW-NB15 \cite{Moustafa2015} datasets. 

\subsection{Federated Learning}

In this case, Nguyen et al. \cite{Nguyen2019} proposed ``DÏoT: A Federated Self-learning Anomaly Detection System for IoT''. DÏoT is a distributed self-learning framework based on FL to monitor IoT devices, composed of two components: (i) Security Gateways, which identify device types and perform local training, and (ii) the IoT Security Service, which maintains device-type-specific Gated Recurrent Unit (GRU) models and coordinates global training. Anomalies are detected as deviations from expected packet behavior. Experiments demonstrated that DÏoT achieves high true positive rates with near-zero false positives.

Next, Li et al. \cite{Li2021} proposed an FL approach where multiple Cyber Physical System (CPS) owners collaboratively train a CNN–GRU model to address two challenges: the scarcity of attack examples for each owner and their reluctance to share sensitive data. Each CPS trains locally, encrypts its model parameters, and sends them to a cloud server for aggregation. Therefore, the contribution ratio of each CPS to the global model is based on the size of its data resource. Evaluated on a real data resource of a gas pipelining system \cite{Morris2014}, the method outperformed state-of-the-art approaches, surpassed locally trained models with limited data, and closely matched a centrally trained “ideal” model with unlimited data, both in standard metrics and in detecting diverse cyber threats. 

Furthermore, Mothukuri et al. \cite{Mothukuri2022} adopted FL to mitigate the risks of raw data transfer and the high cost of centralized training. Seven GRU models with varying window sizes were initialized in a central server, distributed to IoT devices for local training, and aggregated for global updates. After several epochs, the ensemble models were built using Random Forest (a parallel training method). Experiments, using virtualization on a single machine with separate processors for each IoT device, tested four GRU architectures on the Modbus-based dataset \cite{Frazao2019}. It was found that FL consistently outperformed non-FL across performance metrics, window sizes, and GRU architectures—except for a dropout-based one, while reducing training time. However, these results remain questionable due to possible overfitting (no training/testing split details and unexplained discrepancies in dropout performance), and the omission of communication costs, which undermines the claimed efficiency. 

Additionally, Huang et al. \cite{Huang2023} proposed a ``dual
Execution \& Evaluation network FL framework (EEFED)'' framework consisting of two networks: the execution network, which is like a regular FL network, responsible for local training, aggregation, and locally updating model parameters; and the evaluation network responsible for considering data imbalance and the difference in computing power for global model updates and personalized local updates, and regulating the model parameters updating through Reinforcement Learning to assure the stability of the system and minimize fluctuations. Evaluated on the Secure Water Treatment (SWaT) \cite{Mathur2016} and Water Distribution (WaDi) \cite{Ahmed2017} datasets, EEFED achieved higher accuracy and better detection of unknown attacks than baseline methods, with only a small increase in computation. 

\subsection{Generative AI}

In the case of Generative AI, Li et al. \cite{Li2019} proposed ``MAD-GAN:  Multivariate anomaly detection for time series data with generative adversarial networks''; for anomaly detection using Generative Adversarial Networks (GANs). In their approach, a generator is trained to produce fake data resembling real sensor data, while a discriminator learns to distinguish real samples from generated ones. Once trained, the discriminator can detect anomalies directly via the "Discrimination Loss", and the generator via a sample's "Reconstruction Loss". These two components are combined into a unified "DR-Score". Both the generator and discriminator use Long Short-Term Memory (LSTM)-Recurrent Neural Networks. Experiments on the SWaT and WaDI datasets showed that MAD-GAN outperformed other methods. Precision is slightly low, likely due to data imbalance and the system prioritizing high recall to detect all attacks, tolerating some false alarms.

On the other hand, Nie et al. \cite{Nie2022} proposed an IDS for ``CEC-based SIoT systems'' using GANs to address data scarcity. Detection proceeds in two stages. Stage 1: for each attack type, a generator synthesizes samples, and the discriminator decides whether an input (real or synthetic) belongs to that attack type. Stage 2: outputs from all single-attack models are fed to a new generator; a new discriminator is trained to classify inputs as attack or benign. Whether on Stage 1 or 2, both networks are Multi Layer Perceptrons (MLP) and training alternates based on the discriminator’s accuracy on real vs. synthetic data (higher on real → improve the generator; higher on synthetic → improve the discriminator). On CSE-CIC-IDS2018 and CIC-DDoS2019 \cite{Sharafaldin2019}, the method outperformed deep learning baselines and detected previously unseen attacks.

Similarly, Park et al. \cite{Park2023} proposed using GANs to generate synthetic network traffic and mitigate malicious data scarcity in IDS datasets. They employed a five-layer autoencoder discriminator and a generator mirroring the discriminator’s decoder. After training the autoencoder on real and synthetic data, only the encoder part is retained for dimensionality reduction and feature extraction. The final classifier is either the trained encoder followed by a DNN or CNN, or a standalone LSTM. Evaluated on NSL-KDD, UNSW-NB15, IoT-23 \cite{sebastian_garcia_2020}, and a real enterprise dataset, the method improved minority-class detection and outperformed baselines.

\subsection{NLP Methods \& Genetic Algorithms}

In this case, we consider NLP methods and genetic algorithms. First, Deng and Hooi \cite{Deng2021} proposed the Graph Deviation Network (GDN) for anomaly detection in IoT sensor networks. Where a learned embedding represents each sensor. However, cosine similarity between embeddings is used to identify the most influential sensors for modeling another sensor's behavior. These relationships are encoded as directed edges in a graph. For forecasting, an aggregated representation for each sensor combines its embedding and past values with those of its neighbors, with an attention mechanism weighting neighbor contributions by relevance. This aggregated vector is passed to a stacked fully connected network to predict future sensor values, and anomalies are identified as deviations from these predictions. GDN was evaluated on the SWaT and WaDI datasets. 

Next, Wu et al. \cite{Wu2022} proposed ``RTIDS (Robust Transformer-based Intrusion Detection System)'', which leverages a Transformer architecture for three main reasons: (i) its ability to process sequential data using positional encoding and attention for contextual information; (ii) its computational efficiency on network traffic compared to other sequential models (e.g., LSTM, GRU) thanks to parallelization; and (iii) its capacity to compress datasets through embeddings while preserving essential information. RTIDS was evaluated on the CICIDS-2017 and CIC-DDoS2019 datasets.

Moreover, Zhou et al. \cite{Zhou2025} proposed ``HIDIM (Hierarchical Dependency and Class Imbalance)'', a framework composed of two modules: (i) the Hierarchical Semantic Attribute Embedding model, which uses flow IDs and protocol layer IDs to capture hierarchical dependencies across Open Systems Interconnection layers and intra-layer dependencies, respectively. These are integrated into a new flow representation, with an attention mechanism applied to account for differences in feature importance; (ii) the Enhanced Boundary-Oriented Oversampling (EBOO) method, designed for synthetic oversampling and addressing class imbalance with small disjuncts. 

Therefore, in EBOO, borderline samples are identified through a mutual nearest-neighbor approach, weighted according to their proximity to the decision boundary and the sparsity of their cluster—aiming to reinforce weak points of the boundary—and then used to generate minority samples via a Genetic Algorithm (GA), which combines two selected borderline samples through crossover for important features and mutation for less important ones. For final classification, the authors employed a simple MLP with two hidden layers. HIDIM was evaluated on NSL-KDD, UNSW-NB15, AWID2 (Aegean WiFi Intrusion Dataset) \cite{Kolias2016}, CICIDS2017, and NF-BoT-IoT-v2 \cite{NF-BoT-IoT-v2}, where it consistently outperformed baseline methods in accuracy, F1-score, False Negative Rate (FNR), and False Positive Rate, with a notable advantage in reducing FNR.

Finally, Saheed et al. \cite{Saheed2025} proposed the GA-mADAM-IIoT framework for intrusion detection in Industrial IoT (IIoT) environments. The framework combines three key components: (i) an attention mechanism integrated into LSTM to help the model focus on the most relevant timestamps; (ii) a Genetic Algorithm for feature selection, in which feature sets are iteratively combined via crossover and mutation and evaluated using a fitness function that accounts for anomaly detection relevance, dimensionality reduction, and stability across multiple datasets; and (iii) a modified ADAM (mADAM) optimizer that improves momentum updates for sharp or noisy loss landscapes by computing gradients at a look-ahead position, while also adapting the learning rate for each parameter individually based on gradient frequency and magnitude to mitigate vanishing and exploding gradients. To enhance interpretability, the authors employed SHAP for local explainability. GA-mADAM-IIoT was evaluated on the SWaT and WaDI datasets. 

\section{Discussion \& Conclusion}
\label{sec:conc}

To conclude, we examined in this work current trends in applying Artificial Intelligence (AI) to cybersecurity, with an emphasis on intrusion detection. The selected articles/papers for the review were all Scopus-indexed and were selected based on their relevance, recency, and citation count. 

Indeed, we found several emerging trends in our review, including the growing usage of "SHapley Additive exPlanations" for explainability, personalizing local models and contribution ratios to account for heterogeneity among nodes in federated learning, the evolving use of generative adversarial networks for data augmentation and generating synthetic network traffic, and the versatility of natural language processing methods in building Intrusion Detection System (IDS) models due to their adaptability to different entities and contexts. 

However, we also observed notable limitations in many works, such as the neglect of network traffic specificities, insufficient justification for architectural choices, and a lack of attention to explainability and to interpreting the learned patterns and attack behaviors.

Moreover, in the extended version of this conference paper, we aim to include a taxonomy of AI and more eXplainable AI methods, explore additional emerging trends such as the use of Digital Twins and Agentic AI in IDS development, and include more relevant works with a more detailed comparative analysis.

\begin{credits}
\subsubsection*{\ackname} This research study is currently being funded by the Moroccan National Center for Scientific and Technical Research (CNRST: Centre National pour la Recherche Scientifique et Technique) via the PhD-Associate Scholarship – PASS under the grant number 73UAE2024. This content solely represents the opinions of the authors, and the CNRST disclaims any liability for any use of the data/information it includes.

\subsubsection*{\discintname}
Regarding this paper's content, the authors have no conflicting interests to disclose. 
\end{credits}

%
%
\bibliographystyle{unsrt}
\bibliography{references}
%
\end{document}